# Random Forest classifier for EEG-based seizure prediction

Remy Ben Messaoud & Mario Chavez

*Abstract*—Epileptic seizure prediction has gained considerable interest in the computational Epilepsy research community. This paper presents a Machine Learning based method for epileptic seizure prediction which outperforms state-of-the art methods. We compute a probability for a given epoch, of being pre-ictal against interictal using the Random Forest classifier and introduce new concepts to enhance the robustness of the algorithm to false alarms. We assessed our method on 20 patients of the benchmark scalp EEG CHB-MIT dataset for a seizure prediction horizon (SPH) of 5 minutes and a seizure occurrence period (SOP) of 30 minutes. Our approach achieves a sensitivity of 82.07 % and a low false positive rate (FPR) of 0.0799 /h. We also tested our approach on intracranial EEG recordings.

*Index Terms*—Epilepsy, seizure prediction, electroencephalogram, Machine Learning, signal processing, Random Forest.

## I. Introduction

### A. Epileptic seizure prediction

**E**pilepsy is a neurological disorder that affects 1% of the population world-wide [1]. The factor that certainly impacts the most the lives of epileptic patients is the apparently random occurrence of seizures. When seizures occur, the patients generally experience muscle stiffness, spasms or partial or complete loss of consciousness. Thus, epilepsy increases the risks of injuries, fractures or vehicle accidents [2] in addition to the severe psychological impacts on the patients like anxiety, depression or suicide [3]. The identification of early changes in electroencephalographic (EEG) dynamics prior to seizures could be of great interest for patients affected by pharmaco-refractory epilepsy. An efficient prediction system that raises an alarm early enough might allow intervention procedures like drugs or electrical stimulation, or even just minimum safety procedures.

Efforts in computational epilepsy research have been multiplied during the last decade and translated from seizure detection to seizure prediction. Most of these studies assume the hypothesis of the existence of a pre-ictal state. This state would be a transition state from the inter-ictal (resting or normal state) to the ictal state (during which a seizure occurs). The existence of such a state remains questionable until today but many studies have succeeded to show pre-ictal modifications on the EEG dynamics prior to seizures.

### B. Literature review

Important issues regarding the large number of studies in this field are the reproducibility and the comparability. The research community needed a common and solid framework that would allow to assess statistical validity of the results and to compare different studies. In [4][5] authors developed the basis of such a framework by defining two key concepts: the seizure prediction horizon (SPH), which is the minimum time interval separating an alarm to the onset of the seizure, also called intervention time (IT), and the seizure occurrence period (SOP) which is the time interval following the SPH and during which the seizure begins. An important point to keep in mind when comparing studies results is that performances are comparable only if they are measured for the same SPH and SOP, and on the same database. Given the large variability of the choices for these parameters in the literature it is often hard to assess and compare studies results. And yet, some very recent studies [6][7][8][9][10], seem to ignore this framework and only present their classification accuracies [8][9] as a proof of the superiority of their method. Thus, even if these studies show outstanding results, *e.g.* [10] claims to achieve a sensitivity of 99.6% and FPR of 0.004 /h, they are questionable because contrarily to what [10] advances, seizure prediction task cannot be formulated only as a classification task between preictal and interictal brain states. In fact, in these studies [6][8][9][10] authors just performed classification between two segment classes that they called preictal (within one-hour preceding seizures) and interictal (at least four hours before or after any seizure). This methodology was first suggested by the American Epilepsy Society Seizure Prediction Challenge (Kaggle)[11] in the context of a classification competition.

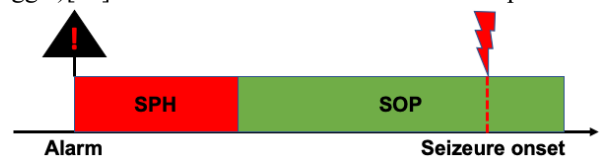

**Figure 1.** Definition of the seizure prediction horizon (SPH) and the seizure occurrence period (SOP). Here is illustrated an example of correct prediction.

Moreover, because this methodology does not include the entire recordings and do not account for the temporality of the signal and the raised alarms, it makes it obsolete for developing seizure prediction systems. That is why we adopted in this paper an alarm system like approach as in [12], that we consider as the most rigorous and complete study in the field so far. Most differences in seizure anticipation algorithms are visible in the two main steps which are feature extraction and classification of preictal against interictal classes. Most recent studies tried to improve this first classification step by translating from Machine Learning techniques to Deep Learning and tried to apply convolutional or recurrent neural networks to the problem. Using artificial neural networks often means the automatization of the first step, feature extraction, [10] or its restriction to the calculation of the spectrogram [8][12][13][14]. Such algorithms can achieve very good results: in [12] for instance, the proposed algorithm achieves a mean sensitivity of 81.2 % and FPR of 0.16 /h for a SPH and SOP of 5 and 30 minutes. the authors of [15], the last review on the subject, consider that [14] performed better by achieving 87.8% of sensitivity and FPR of 0.142/h with a SPH of 10 minutes. That is not completely true because, realistically, the authors of [14] used an implicit SPH of zero and a SOP of 10 minutes but most importantly their study presents serious methodological limits since they did not asses their method with a Leave One Out Cross Validation (LOOCV) and manually selected, instead, a given test set with only 18 seizures. Moreover, the critical issue with deep learning techniques is the interpretability. They are black box models [15] and for that reason they are not easily accepted by the medical community. In addition, their complexity and training time make them less preferable against a simpler algorithm with equal performances. Hence, feature extraction step is a central issue in seizure prediction studies. In [7], authors apply a recurrent neural network (Long Short-Term Memory neural network) on a set of extracted features. They claim to predict all of the 185 seizures of the CHB-MIT with a maximum false positive rate of 0.11/h. That is a very strong result to announce, too good to be true, and usually, in such cases, methodological biases are present. In fact, in the case of [7], beyond the fact that the authors do not use the seizure characteristic evaluation framework [4][5], the problem lies, as for [14], in their way of testing their algorithm. They used a stratified 10-fold cross validation strategy during which they shuffle the segments. Thus, they mix the seizures and do not treat them as independent events. Such a testing method could be seen as equivalent to the Kaggle methodology [11] explained earlier and does not reflect the generalization capacity of the proposed algorithm. A clear limitation, that the authors discussed, is the fact that they could not estimate the prediction time: time laps between the alarm raising and the seizure onset, that is a crucial performance descriptor in seizure prediction studies. Furthermore, except [12][14] the large majority of the studies used SOPs exceeding 40 minutes when [4] demonstrated that, for SOPs exceeding 36 minutes, the sensitivity of the method is no longer a performance feature and that such a waiting time would considerably increase patients stress.

In previous studies, a large set of EEG features have been tested for their seizure predictability performance. Ref. [16] summarizes 24 types of features that could anticipate seizures accurately partitioned into 4 families: statistical, fractal, entropy and spectral. All these features are univariate (estimated for each channel separately). In our work we use a subset of these features which were singled out following the criteria of their ability to discriminate the preictal against the interictal epochs and their computation time. Other studies showed the added value of using bivariate features [17], such as linear correlation, nonlinear interdependence, wavelet synchrony [18], space-delay correlation [19] or phase correlation [20]. Here, we decided to incorporate the linear Pearson's correlation between every pair of channels in our set of features. Overall, the innovative part of this paper lies in the introduction of new postprocessing methods and the exploration of artifacts control, that permit to reduce considerably the false alarms.

The rest of the paper is organized as follows. We present the methods in the first section including data segmentation and labeling, pre-processing, features extraction and selection, classification and pre-ictal alarm detection. We then expose the performances evaluation and statistical validation of our approach. We finally discuss our results and provide some considerations about future work.

## II. METHODS

### A. Datasets

#### CHB-MIT dataset

We applied our method to the CHB-MIT dataset [21]. It consists of continuous scalp EEG recordings from 24 pediatric subjects with intractable seizures who were undergoing withdrawal of antiseizure medication to characterize their seizures for epilepsy surgery evaluation. The recordings were all sampled at 256 Hz and acquired using the 10-20 bipolar setting with 18 to 21 channels. Overall, the dataset contains 916 hours of annotated recordings and 163 seizures. For the sake of the seizure prediction task, seizures which were too close (< SOP) were considered as one unique seizure that begins at the onset of the leading seizure. Concerning the inclusion/exclusion criteria of the individuals, only two studies kept the full cohort of 24 cases so far [7][22]. Ref. [12] rejected half of the cohort considering that keeping subjects having seizures every 2 hours on average was not very critical. Another recent study [10] kept only 8 patients because they took as criterion for interictal segments to be distant from seizures onsets or ends by at least four hours which corresponds to the Kaggle competition configuration [11]. These criteria might be too restrictive and lead to methodology biases and thus, we wanted to keep all the patients for our analysis. Though, some patients had their montages changed through the monitoring. Patient 12 *e.g.* had changes from bipolar to a common reference settings and others (Patients: 13, 15, and 16) do not have a consistent set of electrodes. Hence, we excluded Patients 12, 13 ,15 and 16 from further analysis since our method relies on the consistency of the used montage. This leaves a cohort of 20 patients with 103 seizures and 794.59 interictal hours.

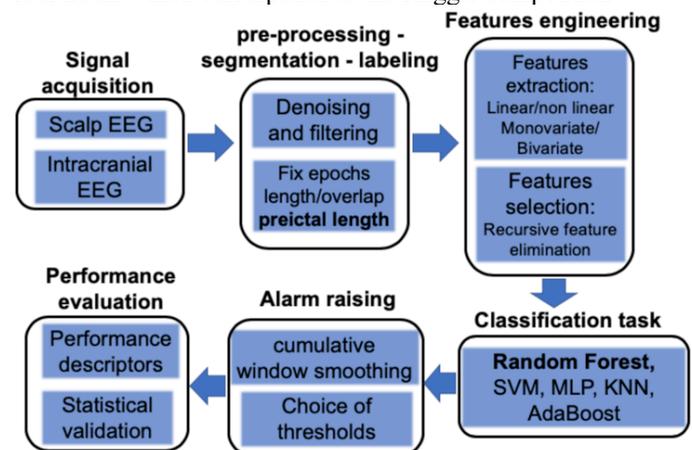

**Fig. 2.** Block diagram of the proposed method.

*Kaggle dataset*

The Kaggle dataset of the American Epilepsy Society Seizure Prediction Challenge [11] consists of intracranial (iEEG) recordings on 7 epileptic cases: 5 dogs and 2 humans. The dataset is already segmented into preictal and interictal periods. Preictal ones are one-hour long recordings beginning 1h:5min and ending 5 minutes before seizures onsets and interictal ones are spaced by at least one week, for dogs, of any seizure and 4 hours for humans. It contains 48 seizures and 627.6 interictal hours. For these recordings, iEEG data was recorded from 16 implanted electrodes with a sampling rate of 400 Hz and 5 kHz for humans, recorded from 15 depth electrodes for Patient1 and 24 subdural electrodes for Patient2 [23].

### B. Seizure prediction approach

The block diagram of the proposed method is illustrated in Fig. 2. It consists of 5 main stages: pre-processing, feature extraction, classification, post-processing, and performance evaluation. A subsection is also dedicated to present the training and testing method.

### C. Pre-processing: data segmentation and filtering

In our study, every EEG file was filtered using the MNE module of Python [24] between 1 and 45 Hz with a FIR 'hamming' window. After estimating the mean and variance of each segment, data were normalized (zero mean and unitary variance). For the human database, most patients had montages with 21 electrodes (if we exclude 2 pairs of electrodes which were collinear: T8-P8 was redundant and T7-P7 had its exact opposite P7-T7), although for patients 17, 18, and 19, one file contained only 18 recorded channels. We interpolated, in that case, the 3 missing channels with MNE which uses the Spherical Splines method [25]. This operation didn't insert any additional information in the dataset and enabled us to have the same number of channels for all files and all patients. Then, data was segmented in epochs of 15 seconds with 2/3 overlap, leading to a step of 5 seconds between the epochs.

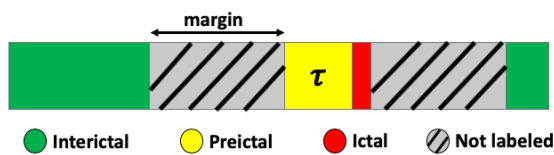

**Fig. 3.** Sketch of the labeling strategy. τ is the preictal length. Margins are located right after seizures end and just before the preictal time. Segments coming from these periods are not considered during the training because of their ambiguity but they are still part of the testing.

### D. Data labeling

This point is very important because it determines the classification performance. In fact, a major step in the seizure prediction is to discriminate preictal against interictal segments. This brings the concept of preictal period/time, time laps preceding seizures, and the question of determining its length. Ref. [14] addresses the problem with a computational approach. They consider the preictal length as a hyperparameter of their algorithm and optimize it through a grid search. They concluded that 10 minutes was the optimum and that it is a sort of inflexion point between interictal and preictal segments. We, therefore, used a preictal length of 10 minutes for the rest of the analysis. We also tried our methodology with a preictal length of 5 and 15 minutes and results showed a clear loss of performance. Additionally, a margin of 20 minutes was taken when labeling data to avoid contaminating the interictal dataset in cases where preictal segments would lie outside the preictal period. Finally, to ensure classes balance, we perform under sampling on the interictal class further explained in subsection *I*. The labeling scheme is summarized in Fig. 3.

### E. Features extraction and selection

At first, we extracted the set of features promoted by [16] which includes 24 types of features. The most sophisticated of these were computed using the PyEEG module of python [26]. We added to that set the largest Lyapunov exponent since [27] had proven its applicability for seizure anticipation. In addition, we also computed the Pearson's correlation coefficients between all the pairs of electrodes. Correlation was computed on the filtered signals (1 - 45 Hz). Overall, we had initially 37 univariate feature types and the correlation between each electrode pair for bivariate features. Therefore, we had a feature space of dimension almost one thousand and it was critical to reduce it. We used for that two criteria: the feature importance and the computation time. The feature importance was calculated using the Recursive Feature Elimination algorithm of Scikit-learn [28] and then averaged over all the patients. Considering these two criteria we kept only 25 monovarietal

**Table 1.** Summary of the feature types extracted for each 15-s epoch. Column FpC gives the number of features extracted per channel.

| Family | Type | Description | FpC |
|---|---|---|---|
| | Mean | Average | 1 |
| | Crest | Maximum value | 1 |
| | Trough | Minimum value | 1 |
| | Var | Variance | 1 |
| Statistical | Skw | Skewness | 1 |
| | Kurt | Kurtosis | 1 |
| | DFA | Detrended Flucyuation Analysis | 1 |
| | Hmob | Hjorth Parameters: Mobility | 1 |
| | Hcomp | Hjorth Parameters: Complexity | 1 |
| | dCorrTime | Decorrelation Time * | 1 |
| Fractal | PFD | Petrosian Fractal Dimension | 1 |
| | HFD10 | Higuchi Fractal Dimension for k=10 | 1 |
| Entropy | SpEn | Spectral Entropy | 1 |
| Spectral | PSI | Power Spectral Intensity ** | 6 |
| | RIR | Relative Intensity Ratio** | 6 |
| Connectivity | corr | Pearson Correlation | 10 |

*first Tau where A(Tau) < 1/e, where A is the Autocorrelation function
**The bands are: delta (0.5 - 4 Hz), theta (4 - 8 Hz), alfa (8 - 13 Hz), beta2 (13 - 20 Hz), beta1 (20 - 30 Hz), gamma (30 - 60 Hz)

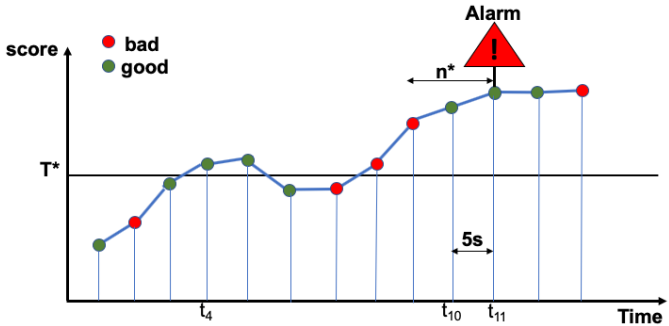

**Fig. 4.** Example of application of the rules of sustainability (n*=3) and artifact rejection with a "good" epochs proportion of 0.5. Because of the sustainability rule, the alarm is not raised at the 4th epoch. The artifact control rule postpones this latter from the 10th to the 11th epoch.

feature types plus the correlation values which leads to a feature space of dimension 735. The retained feature types are summarized in Table. 1.

Furthermore, the prior studies on seizures prediction did not include any form of artifact rejection or control [15]. Therefore, we decided to use a simple Artifact Rejection technique: "Thresholding". The principle is to consider the epochs for which the peak value exceeds a certain threshold as "bad epochs". We introduced, in section II.G a rule-based algorithm to treat these contaminated epochs and we exclude them from training sets. This technique introduces two concepts: the peak value and the threshold. The peak value is already imbedded in our features set since it is the maximum, over all channels, between the magnitudes of the "Crest" and the "Trough". Meanwhile, the threshold is less trivial to define and there is no consensus about it. We decided, to take advantage of the recent efforts of [29] who developed an automated artifact rejection method for EEG and MEG data called *Autoreject* that was integrated into the open source MNE toolbox. This algorithm includes two methods which respectively measure: a global (all channels confused) and a local (per channel) threshold. We used the simplest one (global), and measured one threshold for each file separately (files were 1, 2, or 4 hours long). Each *Autoreject* threshold was then stored and reused during the alarm raising step described further.

### F. Classification step

We decided, for this step, to test the benchmark Random Forest (RF) algorithm [30]. It relies on the aggregation of several decision trees. Ref. [31] demonstrated its efficacy in the context of seizure detection and [9] reported successful results (85.2% of accuracy) for seizure prediction but only in a classification framework and not with the seizure characteristics framework[4]. We also tested the benchmark Support Vector Machine classifier (SVC) given that a large number of studies used this latter [32][33][34]. In addition, we also tested other popular classifiers as k-Nearest Neighbors (KNN), Multilayer Perceptron (MLP) and the AdaBoost classifier. Several output measures were considered. The most trivial is the label, "0" or "1" for the two classes respectively: "interictal" and "preictal". The second measure considered was the probability, for a given epoch, of being in the class "preictal". And the last measure, called here "contrast" and detailed in (1), is a sort of scaling of the second and gave the best performances.

$$\text{"contrast"} = P(\text{"preictal"}) - P(\text{"interictal"}) = 2 * P(\text{"preictal"}) - 1 \quad (1)$$

After a grid search, the retained parameter values are: n=107 for number of estimators of the RF, C=40 for the regularization parameter of the SVC, k=6 for the number of nearest neighbors, alpha = 0.0002 for the regularization of the MLP and n=70 for the number of estimators of the AdaBoost classifier.

### G. Postprocessing and alarm raising

This step consists in the summation of the outputs returned by the classification task within a temporal window. For example, if we choose a length of 10 consecutive epochs, the cumulative score will be defined, respectively for the different measures: "labels", "preictal probability" and "contrast", on {0,1,2...10}, [0,10], and [-10,10]. We then lowpass filter the obtained signal to denoise it (Hamming FIR window: order = 20; cutoff = 0,03*Nyquist frequency= 1,87 Hz). The idea of integrating past values and low pass filtering was inspired by [35] in which authors demonstrated the utility of Kalman filtering and Firing power methods applied to the real valued classification output. Then, the decision rule depends on two parameters: the score threshold, T*, and the sustainability or temporal threshold, n*. An alarm is raised if the score is greater or equal to T* during n* consecutive epochs. Introducing this sustainability threshold is a novelty in the field of seizure anticipation and it permits to considerably reduce the number of false alarms. Moreover, an additional novelty is the artifact control. Each epoch will be labeled as "bad", which means is probably an artifact, if its peak value exceeds the *Autoreject* threshold. This threshold is obtained following the training and testing method explained further. Then, we decide of the proportion of "good" epochs necessary to raise an alarm. An example of this procedure is illustrated in Fig. 4 for n* = 3 and a "good" epochs proportion of $\beta_{good}$ =0.5.

### H. Seizure prediction characteristics: SOP-SPH

Ref. [4] introduced two key quantities: SPH and SOP depicted in Fig. 1. The seizure prediction horizon (SPH) is the time directly following an alarm raise during which the patients can take precautions to avoid the seizure by taking certain medications or just to make sure to be in safe conditions. For this reason, SPH is often called Intervention time (IT) in literature and should be sufficiently long to enable that (at least 2-3 minutes). The seizure occurrence period begins right after the SPH and represent the time period during which seizure onset could happen at any moment. This period is stressful for the patient because he is constantly expecting the seizure to begin. Therefore, seizure prediction methods should try to reduce this quantity. The following rules are considered when evaluating a potential alarm:
  i. An alarm is true if seizure onset happens during the SOP.
  ii. An alarm is false if seizure onset happens during the SPH.
  iii. A new alarm can be raised only after the end of SOP or the occurrence of a seizure.

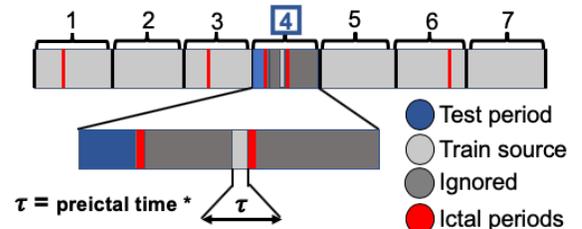

**Fig. 5.** Illustration of the data splitting into train and test. In this case, we test the first subpart of the 4th period. Each time, training data is picked from the train source as follows: we take all the available preictal segments then we randomly pick the same number of interictal segments. *Preictal time is fixed to 10 minutes in the whole analysis. All epochs in these periods are labeled as pre-ictal.

### I. Training and testing method

Almost all recent studies that have been conducted on seizure prediction used the Leave-one-out cross validation (LOOCV). The concept is, for a given patient who made N seizures, to consider all the seizures as training data except one that is used for the testing. This operation is repeated N times with a rollover on the tested seizure. We also used LOOCV but on the full recorded period. It is a sort of generalization of the LOOCV since the seizures are treated the same way but it also achieves the same goals for the interictal periods. In the following example, illustrated in Fig. 5: a patient made 5 seizures during 28 hours of recording, this gives 7 separate periods of 4 hours. Our method is equivalent to a 5-folds-LOOCV on the seizures coupled with a 7-folds-LOOCV on the periods. The result of this operation is a vector of scores, given by the postprocessing step, that describes the entire recorded duration. In addition, the *Autoreject* threshold that we apply on the test period is obtained by averaging the thresholds learned from the rest of the periods. We apply this method on 5 independent runs for each patient to get a mean and a standard deviation of the performance criteria described in the next section.

Most importantly, this way of splitting data ensures that we would almost never have a test epoch and a train epoch that are very close in time. This is very important because of the non-stationarities of EEG signals.

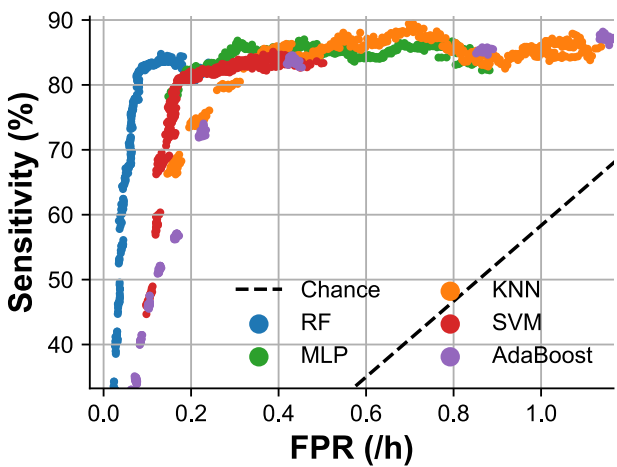

**Fig. 6.** Scatter plot of FPR and Sensitivity for obtained with 5 classifiers: Random Forest, Multi-Layer Perceptron, K-Nearest Neighbors, Super Vector Machine, and AdaBoost. It is not a ROC curve because several parameters have been varied, not only the amplitude threshold, to obtain it. It clearly shows that RF generates very few false alarms.

### J. Performance evaluation

We used, to assess the performance of our methodology, the seizure prediction characteristic [4] which introduces two performance descriptors: the Sensitivity and the False Positive Rate (FPR). The sensitivity is the ratio of the number of predicted seizures over the total number of leading seizures and is measured in (%). The FPR is the total number of false alarms divided by the length of interictal periods and is measured in (/h). What is called interictal periods is dependent on the prior choice of SOP and SPH and it is all the periods preceding a leading seizure by at least SOP + SPH since in this latter interval, by definition, no false alarm can happen. Some studies used less strict definitions of FPR that inevitably lead to lower values. In [12], authors just consider the ratio of false alarms over the total recording's length. In [22], a postictal time is introduced (directly following seizure end) and subtracted from interictal periods. This latter study also measures another specificity descriptor. Its computation is detailed in (2), where the false waiting time is the number of false alarms multiplied by SPH+SOP.

$$specificity = 1 - false\ waiting\ time/interictal\ time$$
$$= 1 - FPR \cdot (SPH + SOP) \quad (2)$$

It is also of common usage to compare the prediction performances to that of an unspecific random predictor. We used for this the same method as [12] that considers a random predictor for which all seizure's events are independent and that the generation of an alarm follows a Poisson's law. Therefore, the probability of raising an alarm during an SOP given a certain FPR is approximately

$$P = 1 - e^{-FPR \cdot SOP} \quad (3)$$

Consequently, the probability that such a predictor detects correctly at least m out of M seizures is given by

$$p = \sum_{k \geq m} \binom{M}{k} P^k (1-P)^{M-k}. \quad (4)$$

This probability is computed for each subject. if m is the number of seizures that our approach predicted for a given subject and *p(m, M)* is less than 0.05 than we can reject the hypothesis that the performance achieved by our method could be achieved by a random Poisson alarm generator at a significance level of 0.05.

## III. RESULTS

We applied our methodology to the CHB-MIT dataset with all the previously introduced classifiers: KNN, SVM, MLP, AdaBoost, and Random Forest and for each output metric: 'labels', 'preictal probabilities', and 'contrast' like output. We concluded that 'contrast' was the best output metric out of the three and only report the corresponding results. We explained in the literature review why we consider [12] as the most rigorous study so far. Thus, we chose similar seizure characteristics to be able to compare our method to theirs (*i.e.* SPH = 5min and SOP = 30min).

We performed a grid search on the calibration parameters of the algorithm *i.e.* the thresholds and tuning factors introduced in the section II.G. That enabled us to plot an equivalent of a ROC curve that permits to visually compare the performances of each classifier. This ROC like curve is displayed in Fig. 6. Optimal points are then chosen with a rule of thumb for each classifier. Fig. 7 summarizes the retained optimums in terms of sensitivity and specificity. The results show the clear superiority of the Random Forest classifier with a sensitivity of 82.07% and a very low FPR of 0.0799/h (i.e. specificity of 95.33%) with an average prediction time of 13.09 minutes. We outperform [12] which achieves 81.2% sensitivity and 0.16/h FPR on only 13 patients of the CHB-MIT dataset. These retained performances were obtained using the following calibration parameters: number of cumulated segments= 11; T*=6.31; n*=7; AF=1.49; and $\beta_{good}$ =0.5. The detailed results for individual cases are summarized in Table. 2.

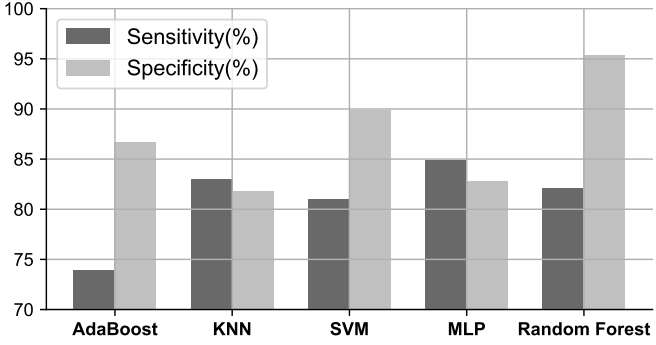

**Fig. 7.** Bar plot of the retained values of sensitivity and specificity for each classifier. The optimums are chosen to be the closest possible to the left corner of the ROC like curve with a certain subjectivity about the weights attributed to sensitivity and specificity.

We also tested the method with the Random forest classifier on the Kaggle dataset. The results are summarized in Table. 3. Contrarily to the CHB-MIT dataset, we calibrated the model for each case separately because the optimization on the population level gave very poor results. We achieve a sensitivity of 69.76% and FPR of 0.214/h (i.e. specificity of 89.91%) with an average prediction time of 16.37 minutes. We do not exceed the scores of [12] since they had a larger sensitivity (75%) for a similar FPR (0.21/h).

**Table 2.** Seizure prediction results obtained on the retained cohort of the CHB-MIT dataset for SPH and SOP of 5 and 30 minutes respectively. These are the optimal performances obtained with the Random Forest classifier. In the #Seizures field, when the number of leading seizures does not correspond to the original number of seizures, this latter is given between parenthesis. For each descriptor, the reported values are the mean and standard deviation computed on five independent runs.

| Cases | # Seizures | Interictal hours | Sensitivity (%) | Specificity (%) | Pred.Time (min) | FPR (/h) | p |
|---|---|---|---|---|---|---|---|
| chb1 | 7 | 36.29 | 97.14 ± 5.71 | 85.825 ± 0.64 | 14.14 ± 1.04 | 0.243 ± 0.011 | 1.2e-07 |
| chb2 | 3 | 33.36 | 66.67 ± 0.00 | 96.15 ± 1.22 | 12.90 ± 4.09 | 0.066 ± 0.022 | 1.1e-04 |
| chb3 | 6 (7) | 34.34 | 100.00 ± 0.00 | 92.88 ± 1.69 | 9.94 ± 0.83 | 0.122 ± 0.029 | 2.4e-08 |
| chb4 | 4 | 153.55 | 95.00 ± 10.00 | 99.53 ± 0.29 | 8.00 ± 0.57 | 0.008 ± 0.005 | 2.1e-10 |
| chb5 | 5 | 35.92 | 96.00 ± 8.00 | 95.1 ± 1.75 | 9.03 ± 1.01 | 0.084 ± 0.030 | 7.8e-08 |
| chb6 | 9 (10) | 61.40 | 68.89 ± 8.31 | 98.83 ± 0.93 | 7.79 ± 0.26 | 0.020 ± 0.016 | 4.8e-17 |
| chb7 | 3 | 65.22 | 40.00 ± 13.33 | 99.82 ± 0.35 | 21.56 ± 1.14 | 0.003 ± 0.001 | 2.3e-08 |
| chb8 | 5 | 17.00 | 80.00 ± 0.00 | 80.80 ± 4.14 | 14.34 ± 3.37 | 0.329 ± 0.071 | 2.3e-04 |
| chb9 | 4 | 65.45 | 95.00 ± 10.00 | 96.96 ± 0.7 | 15.99 ± 1.86 | 0.052 ± 0.012 | 3.7e-07 |
| chb10 | 7 | 45.83 | 100.00 ± 0.00 | 98.71 ± 0.00 | 14.96 ± 0.95 | 0.022 ± 0.000 | 9.7e-15 |
| chb11 | 3 | 32.89 | 60.00 ± 13.33 | 96.09 ± 0.7 | 9.71 ± 4.38 | 0.067 ± 0.012 | 1.2e-04 |
| chb14 | 6 (8) | 22.39 | 70.00 ± 6.67 | 97.90 ± 1.92 | 8.16 ± 0.33 | 0.036 ± 0.033 | 2.8e-10 |
| chb17 | 3 | 19.17 | 60.00 ± 13.33 | 88.45 ± 3.55 | 15.82 ± 6.16 | 0.198 ± 0.061 | 2.6e-03 |
| chb18 | 5 (6) | 32.56 | 76.00 ± 8.00 | 96.79 ± 2.1 | 13.72 ± 0.78 | 0.055 ± 0.036 | 5.9e-08 |
| chb19 | 3 | 28.05 | 66.67 ± 0.00 | 97.90 ± 0.00 | 20.93 ± 0.19 | 0.036 ± 0.000 | 1.8e-05 |
| chb20 | 6 (8) | 23.97 | 100.00 ± 0.00 | 92.70 ± 0.00 | 22.76 ± 0.03 | 0.125 ± 0.000 | 2.2e-08 |
| chb21 | 4 | 30.36 | 100.00 ± 0.00 | 98.83 ± 0.93 | 8.69 ± 1.68 | 0.020 ± 0.016 | 6.9e-09 |
| chb22 | 3 | 29.12 | 100.00 ± 0.00 | 98.42 ± 2.33 | 10.02 ± 1.10 | 0.027 ± 0.040 | 2.1e-06 |
| chb23 | 6 (7) | 23.02 | 90.00 ± 8.16 | 94.92 ± 2.80 | 14.13 ± 2.83 | 0.087 ± 0.048 | 1.8e-08 |
| chb24 | 11 (16) | 4.70 | 80.00 ± 3.64 | 100 ± 0.00 | 9.21 ± 0.62 | 0.00 ± 0.00 | - |
| **Total** | **103** | **794.59** | **82.07 ± 17.33** | **95.33 ± 4.93** | **13.09 ± 4.52** | **0.0799 ± 0.08** | **2.4e-71** |

**Table 3.** Seizure prediction results obtained on the Kaggle dataset for SPH and SOP of 5 and 30 minutes respectively. These are the optimal performances obtained with the Random Forest classifier. The model is calibrated for each case individually.

| Case | # Seizures | Interictal hours | Sensitivity (%) | Specificity (%) | Pred.Time (min) | FPR (/h) | p |
|---|---|---|---|---|---|---|---|
| Dog 1 | 4 | 80 | 91.67 ± 11.79 | 80.81 ± 1.75 | 13.17 ± 1.29 | 0.329 ± 0.03 | 4.78e-04 |
| Dog 2 | 7 | 83.3 | 66.67 ± 6.73 | 95.74 ± 0.29 | 17.01 ± 0.7 | 0.073 ± 0.005 | 1.6e-09 |
| Dog 3 | 12 | 240 | 100.00 ± 0.00 | 89.56 ± 0.7 | 21.21 ± 0.52 | 0.179 ± 0.012 | 1.11e-13 |
| Dog 4 | 14 | 134 | 50.00 ± 0.00 | 100 ± 0.00 | 7.34 ± 0.02 | 0.00 ± 0.00 | - |
| Dog 5 | 5 | 75 | 80.00 ± 0.00 | 95.74 ± 0.35 | 20.30 ± 0.71 | 0.073 ± 0.006 | 2.98e-07 |
| Pat 1 | 3 | 8.3 | 66.67 ± 0.00 | 88.10 ± 0.00 | 18.14 ± 0.10 | 0.204 ± 0.00 | 0.0023 |
| Pat 2 | 3 | 7 | 33.33 ± 0.00 | 79.41 ± 0.00 | 17.42 ± 0.14 | 0.353 ± 0.00 | 0.0176 |
| **Total** | **48** | **627.6** | **69.76 ± 21.5** | **89.91 ± 7.23** | **16.37 ± 4.40** | **0.214 ± 0.124** | **-** |

Once a classifier has been identified as the most suited for the seizure prediction task one can tune the calibration parameters for each patient individually. It is a sort of personalization. We choose, for each single case, from the grid search results, the parameters that give optimal performances. Detailed results are summarized in Table. 5. This process improves significantly the performances leading to an average sensitivity of 89.31% and FPR of 0.03/h. We manage to achieve perfect prediction (*i.e.* predict all seizures with no false alarm) for 4 cases out of 20.

We benchmarked the most recent studies that used the CHB-MIT dataset and summarize the results in Table. 5. Besides the choice of SPH-SOP, regarding the methodology and validation strategy, our results can be directly compared with both [12][22]. The rest of the studies show very good results and are included here mostly to testify of the wide range of choices that can be made in terms of methods and evaluations strategies.

## IV. DISCUSSION

We developed a method that includes both frequency and time domain features but also functional connectivity. It has the advantage that it does not require any further features engineering or electrodes selection that could be a laborious process for experts. This brings the question of the personalization of the algorithm for each patient. In fact, it is an

**Table 4.** Seizure prediction results obtained on the retained cohort of the CHB-MIT dataset for SPH and SOP of 5 and 30 minutes respectively. These are the optimal performances obtained with the Random Forest classifier. The model is calibrated for each case separately. For each descriptor, the reported values are the mean and standard deviation computed on five independent runs.

| Case | Sensitivity (%) | FPR (/h) | Case | Sensitivity (%) | FPR (/h) |
|---|---|---|---|---|---|
| chb1 | 100 ± 0.00 | 0.11 ± 0.024 | chb14 | 76.67 ± 13.33 | 0.018 ± 0.022 |
| chb2 | 66.67 ± 0.00 | 0.012 ± 0.015 | chb17 | 93.33 ± 13.33 | 0.00 ± 0.00 |
| chb3 | 100.00 ± 0.00 | 0.116 ± 0.032 | chb18 | 100.00 ± 0.00 | 0.068 ± 0.03 |
| chb4 | 100.00 ± 0.00 | 0.00 ± 0.00 | chb19 | 66.67 ± 0.00 | 0.00 ± 0.00 |
| chb5 | 100.00 ± 0.00 | 0.106 ± 0.027 | chb20 | 100.00 ± 6.67 | 0.058 ± 0.031 |
| chb6 | 88.89 ± 8.31 | 0.049 ± 0.031 | chb21 | 100.00 ± 0.00 | 0.00 ± 0.00 |
| chb7 | 46.67 ± 16.33 | 0.00 ± 0.00 | chb22 | 100.00 ± 0.00 | 0.00 ± 0.00 |
| chb8 | 100.00 ± 0.00 | 0.012 ± 0.024 | chb23 | 100.00 ± 0.00 | 0.043 ± 0.027 |
| chb9 | 100.00 ± 0.00 | 0.00 ± 0.00 | chb24 | 87.27 ± 4.45 | 0.00 ± 0.00 |
| chb10 | 100.00 ± 0.00 | 0.017 ± 0.009 | | | |
| chb11 | 60.00 ± 13.33 | 0.00 ± 0.00 | **Total** | **89.31 ± 16.21** | **0.03 ± 0.04** |

**Table 5.** Benchmark of recent studies that used the CHB-MIT dataset for the seizure prediction task. The field "same calibration" details whether calibration of alarm raising system was personalized or the same for the whole cohort.

| Year | Authors | Dataset | Features | Classifier | Same FE** | Same calibration | #Seizures | Sensitivity (%) | FPR (/h) | SOP (min) | SPH (min) |
|---|---|---|---|---|---|---|---|---|---|---|---|
| 2016 | Zhang, Z., & Parhi, K. K. [32] | MIT, 17 patients | Power spectral density ratio | SVM | No | No | 80 | 98,68 | 0,05 | 50 | 0* |
| 2017 | Alotaiby, T. N. et al., [22] | MIT, 24 patients | CSP | LDA | No | Yes | 170 | 81<br>87<br>89 | 0,47<br>0,4<br>0,39 | 60<br>90<br>120 | 0 |
| 2018 | Khan, H. et al., [14] | MIT, 15 patients | Wavelet transform | CNN | Yes | Yes | 18 | 83,33 | 0,147 | 10 | 0* |
| 2018 | Truong, N.D. et al., [12] | MIT, 13 patients<br>Kaggle | Short-time Fourier transform | CNN | Yes | Yes | 64<br>48 | 81,2<br>75 | 0,16<br>0,21 | 30 | 5 |
| 2019 | Daoud, H. & Bayoumi, M.A [10] | MIT, 8 patients | DCAE + Bi-LSTM | No | Yes | 43 | 99,6 | 0,004 | 60 | 0* |
| 2020 | **this work** | MIT, 20 patients<br>MIT, 20 patients<br>Kaggle | Table 1 | Random Forest | Yes | Yes<br>No<br>No | 103<br>103<br>48 | 82,07<br>89,31<br>69,76 | 0,0799<br>0,03<br>0,21 | 30 | 5 |

FE, Features Engineering; MIT, Massachusetts Institute of Technology scalp EEG dataset; SVM, Support Vector Machine; CSP, Common Spatial Patterns; CNN, Convolutional Neural Network; LDA, Linear Discriminant Analysis; DCAE, Deep Convolutional Autoencoder; Bi-LSTM, Bidirectional Long Short-Term MemoryRecurrent Neural Network;

\* The authors implicitly used a zero SPH disregarding clinical caracteristics of seizure prediction developed in [4]

\*\* "No" means that Features Engineering was not the same across all patients or in the case of [22] it was handled by a CSP or by an DCAE for [10]

open question and is of great importance in the seizure prediction context. It can considerably improve seizure prediction performances [36] but also adds additional complexity to the proposed methods. One can see the paradigm this way, an epileptic person wants to use a potential seizure prediction alarm system. In all cases, this latter will have to be monitored with video EEG during at least 24 hours and to go through 3 seizures. Then, with a non-personalized method, the patient could directly begin to use the system whereas for a personalized method, the personalization would have to be carried out by an expert and validated.

Personalization can be done in almost all the steps of the seizure prediction algorithm. We identify here four different levels: segmentation and labeling, features and electrodes selection, alarm system calibration, and seizure characteristics (SPH-SOP) choice. The literature on the field contains some personalization attempts. To begin with the first level, [36] showed the impact of personalization of pre-seizures parameters especially the sliding window size and the preictal time (that was fixed to 10 minutes here). Concerning features and electrodes selection, some studies like [32] applied a patient-specific 2-steps features/electrode selection. Besides, we can consider as a form of personalization the fact of delegating the features extraction step to an algorithm. For example, [22] used Common Spatial Patterns for automatic features extraction and [10] used a Recurrent Neural Network coupled with an electrode selection process. In our case, we selected via the Recursive Features Elimination algorithm a given set of features that is kept identical for all patients. On the third level, alarm system calibration, no explicit examples have been reported. Technically, this calibration would usually be performed through a grid search whether on the patient level or on the cohort level. We reported a significant improvement in performances when personalizing the calibration from 82.07% to 89.31% for sensitivity and 0.0799/h to 0.03/h for FPR. Finally, the last level of personalization is the choice of the seizure characteristics SPH-SOP. These are key parameters and are probably the most appropriate to personalize. One attempt has been made by [37] in which authors used the Mamdani fuzzy inference system to find optimal SPH-SOP pairs. They achieved 100% sensitivity and average FPR of 0.13/h on ten patients of the Freiburg dataset [38].

For future work, a first step would be to validate the method on additional data beginning with the benchmark Freiburg dataset [38]. We could also go further in the exploration of personalization effects by performing patient-specific features and electrodes selection for example. And finally, an interesting further progress would be to investigate the feasibility of our method in an online seizure prediction framework. We believe it is feasible because the average processing time of one 15s window ranged from 0.5 to 1 second with python which is fine if the epochs step is fixed to five seconds.

ACKNOWLEDGMENT

This research was supported by the Paris Brain Institute informatics Hub. The authors appreciate Dr. Arthur Tenenhaus from the Signals and Statistics department of the School of Engineering CentraleSupélec, for the discussion about the Random Forest algorithm.